\documentclass[twocolumn,twoside,slac_two]{revtex4}
\usepackage{graphicx}
\usepackage{fancyhdr}
\begin{document}

%Title of paper
\title{Submillimeter Variability and the Gamma-ray Connection in \textit{Fermi} Blazars}

\author{A. Strom}
\affiliation{Univ. of Arizona, AZ 85721, USA}
\author{A. Siemiginowska, M. Gurwell, B. Kelly}
\affiliation{CfA, MA 02138, USA}

\begin{abstract}
We present multi-epoch observations from the \textit{Submillimeter
Array} (\textit{SMA}) for a sample of 171 bright blazars, 43 of which
were detected by \textit{Fermi} during the first three months of
observations. We explore the correlation between their gamma-ray
properties and submillimeter observations of their parsec-scale jets,
with a special emphasis on spectral index in both bands and the
variability of the synchrotron component. Subclass is determined using
a combination of \textit{Fermi} designation and the Candidate
Gamma-Ray Blazar Survey (CGRaBS), resulting in 35 BL Lac objects and
136 flat-spectrum radio quasars (FSRQs) in our total sample. We
calculate submillimeter energy spectral indices using contemporaneous
observations in the 1 mm and 850 micron bands during the months
August--October 2008. The submillimeter light curves are modeled as
first-order continuous autoregressive processes, from which we derive
characteristic timescales. Our blazar sample exhibits no differences
in submillimeter variability amplitude or characteristic timescale as
a function of subclass or luminosity. All of the the light curves are
consistent with being produced by a single process that accounts for
both low and high states, and there is additional evidence that
objects may be transitioning between blazar class during flaring
epochs.

\end{abstract}

\maketitle

\thispagestyle{fancy}

\section{INTRODUCTION}

The timescales on which high-amplitude flaring events occur in blazars
indicate that much of the energy is being produced deep within the jet
on small, sub-parsec scales \citep{sikora01AIPCS, sikora01ASPCS}.
Understanding if/how emission differs between blazar subclasses (i.e., BL Lacs
objects and flat-spectrum radio quasars (FSRQs)) may offer important
insight into the similarity between blazars and, furthermore, can
provide constraints on the formation and acceleration of the jets
themselves.

For the synchrotron component of blazar spectra, the low-frequency
spectral break due to synchrotron self-absorption moves to higher
frequencies as one measures closer to the base of the jet
\citep{sikora01ASPCS}. This often places the peak of the spectrum in the millimeter and submillimeter bands, where the emission is optically-thin and
originates on parsec and sub-parsec scales
\citep{stevens94}, allowing direct observation of the most compact regions near the central engine. The high energy $\gamma$-ray
emission originates as a Compton process, typically a combination of synchrotron-self-Compton (SSC) and external-radiation-Compton (ERC). Depending on the source
properties, the synchrotron photons or external photons are upscattered
by the same population of electrons that emit the millimeter and
submillimeter spectra. Therefore the submillimeter and $\gamma$-ray
emission are closely linked and give the full information about the
source emission.

A systematic study of the submillimeter properties of the entire
sample of \textit{Fermi} blazars has yet to be conducted and is one of
the primary goals of our work. We present here preliminary
analysis of the submillimeter properties of \textit{Fermi} blazars detected by the
\textit{Submillimeter Array}\footnote{The Submillimeter Array is a
joint project between the Smithsonian Astrophysical Observatory and
the Academia Sinica Institute of Astronomy and Astrophysics and is
funded by the Smithsonian Institution and the Academia Sinica.}
(\textit{SMA}) at 1mm and 850$\mu$m, including an investigation of
variable behavior and the determination of submillimeter energy
spectral indices. In addition, we consider the connection to the
observed $\gamma$-ray indices and luminosities.

\section{\textit{SMA} BLAZARS}

The \textit{Submillimeter Array} \citep{ho04} consists of eight 6 m
antennas located near the summit of Mauna Kea.  The \textit{SMA} is
used in a variety of baseline configurations and typically operates in
the 1mm and 850$\mu$m windows, achieving spatial resolution as fine as
0.25'' at 850$\mu$m. The sources used as phase calibrators for the
array are compiled in a database known as the \textit{SMA} Calibrator
List\footnote{http://sma1.sma.hawaii.edu/callist/callist.html}
\citep{gurwell07}.  Essentially a collection of bright objects
(stronger than 750 mJy at 230 GHz and 1 Jy at 345 GHz), these sources
are monitored regularly, both during science observations and
dedicated observing tracks.

To select our sample, we identified objects in the calibrator list
that were also classified as BL Lacs or FSRQs by the Candidate Gamma-Ray Blazar Survey
\citep[CGRaBS]{healey08}.  Of the 243 total
objects in the calibrator list, 171 (35 BL Lacs and 136 FSRQs) have
positive blazar class identifications, although there are three
sources (J0238+166, J0428-379, and J1751+096) which have conflicting
classifications between \textit{Fermi} and CGRaBS.  Some blazars found
in the calibrator list have been studied extensively (e.g., 3C 279 and
3C 454.3) but the \textit{SMA} blazars have not been studied
collectively.

Forty-four of the objects in our total blazar sample were detected by
\textit{Fermi}  and can be found in the catalog of LAT Bright
AGN Sources (LBAS) from Abdo et al. \cite{abdo09}. J0050-094 has no redshift in
either the LBAS catalog or CGRaBS and is not included in our study. Of the 43 remaining sources, 14 are BL Lac
objects and 29 are FSRQs, with $0.03 \leq z \leq 2.19$.

We examined submillimeter light curves for all of the 
\textit{SMA} blazars, with observations beginning in approximately 2003 (see Figure 1). Typically,
the 1mm band is much more well-sampled in comparison to the 850µm
band, but visual inspection reveals that the regularity and quality of
observations vary greatly from source to source. Many of the objects exhibit nonperiodic
variability, either in the form of persistent, low-amplitude
fluctuations or higher amplitude flaring behavior.

\begin{figure}[t]
\begin{center}
\includegraphics[width=0.50\textwidth]{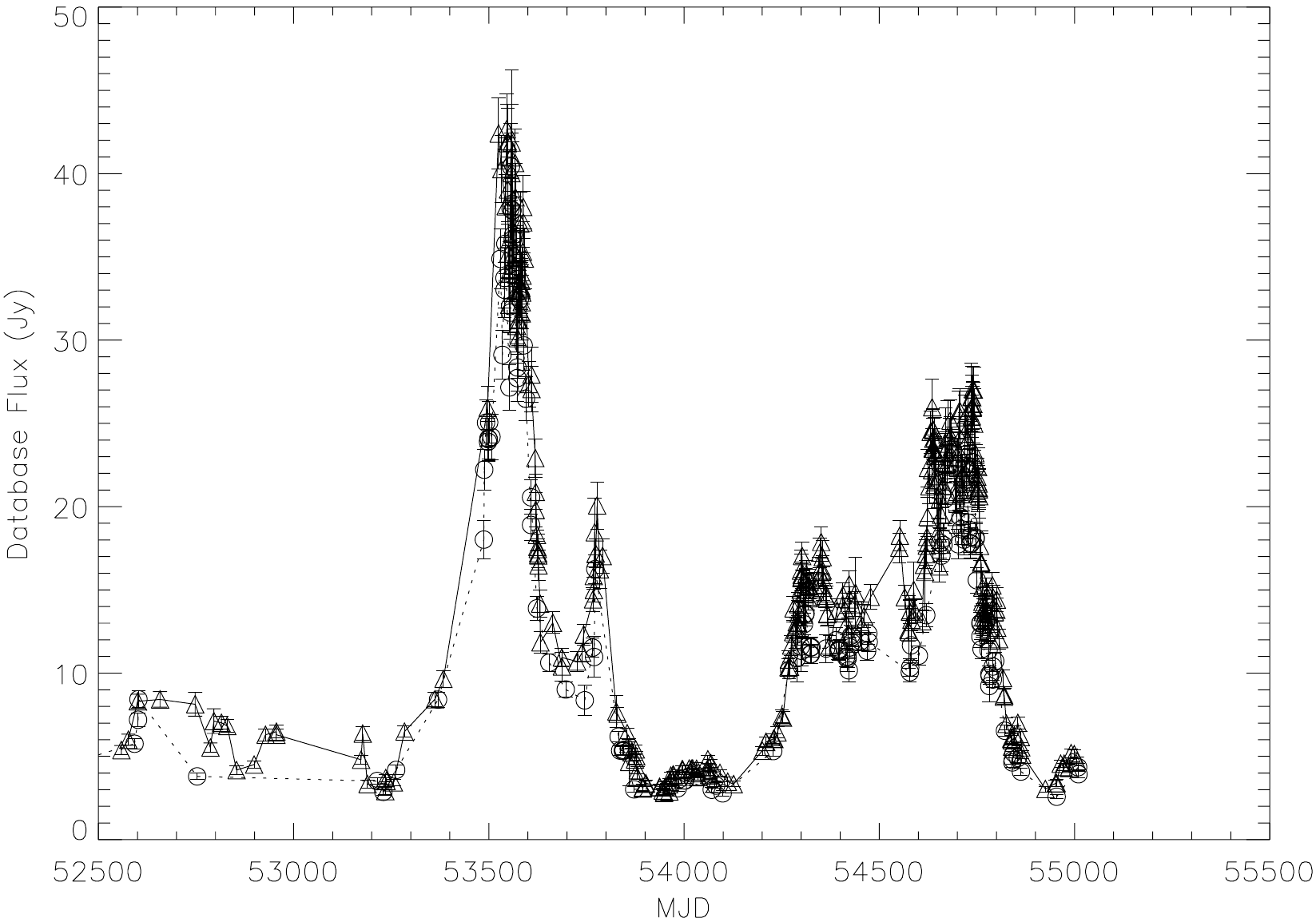}
\caption{The \textit{SMA} light curves for 3C 454.3. The open circles represent the 850$\mu$m observations, and the open triangles represent the 1mm observations.}
\end{center}
\end{figure}

\subsection{Submillimeter Properties}

{\bf Submillimeter Luminosities.} Since we are primarily concerned with comparisons to
\textit{Fermi} observations, we note that only 129 of the \textit{SMA} blazars (23 BL Lacs
and 106 FSRQs) were observed by the \textit{SMA} in either band during the
three months August-October 2008. For these objects, submillimeter
luminosities are calculated in the standard way:
\begin{eqnarray}
\nu_e L_{\nu_e}=4\pi D_{\rm L}^2
   {\nu_{\rm{obs}}F_{\rm{obs}}
\over 1+z},
\label{eq:smalum}
\end{eqnarray}
where $D_{\rm L}$ is the luminosity distance, $\nu_{\rm obs}$ is the frequency of the observed band, and $F_{\rm obs}$ is the
average flux (in erg cm$^{-2}$ s$^{-1}$ Hz$^{-1}$) over the three month period.
We adopt a lambda cold dark
matter cosmology with values of $H_0 = 71$ km
s$^{-1}$ Mpc$^{-1}$, $\Omega_{\mathrm{M}} = 0.27$, and $\Lambda =
0.73$.

\bigskip

{\bf Energy Spectral Indices.} We derive submillimeter spectral energy
indices from observations quasi-simultaneous with the
\textit{Fermi} observations. To be consistent with the use of
$\alpha_\gamma$, we define spectral energy index as $\nu F_\nu =
\nu^{-\alpha_{\mathrm{S}}}$ and calculate $\alpha_{\mathrm{S}}$ from
the average of the energy spectral indices over the corresponding three months. We only
calculate $\alpha_{\rm S}$ for the 16 objects (8 BL Lacs and 35 FSRQs)
with observations at both 1mm and 850$\mu$m during this time frame.
\section{VARIABILITY ANALYSIS}

\subsection{Variability Index}

\begin{figure}[t]
\begin{center}
\includegraphics[width=0.50\textwidth]{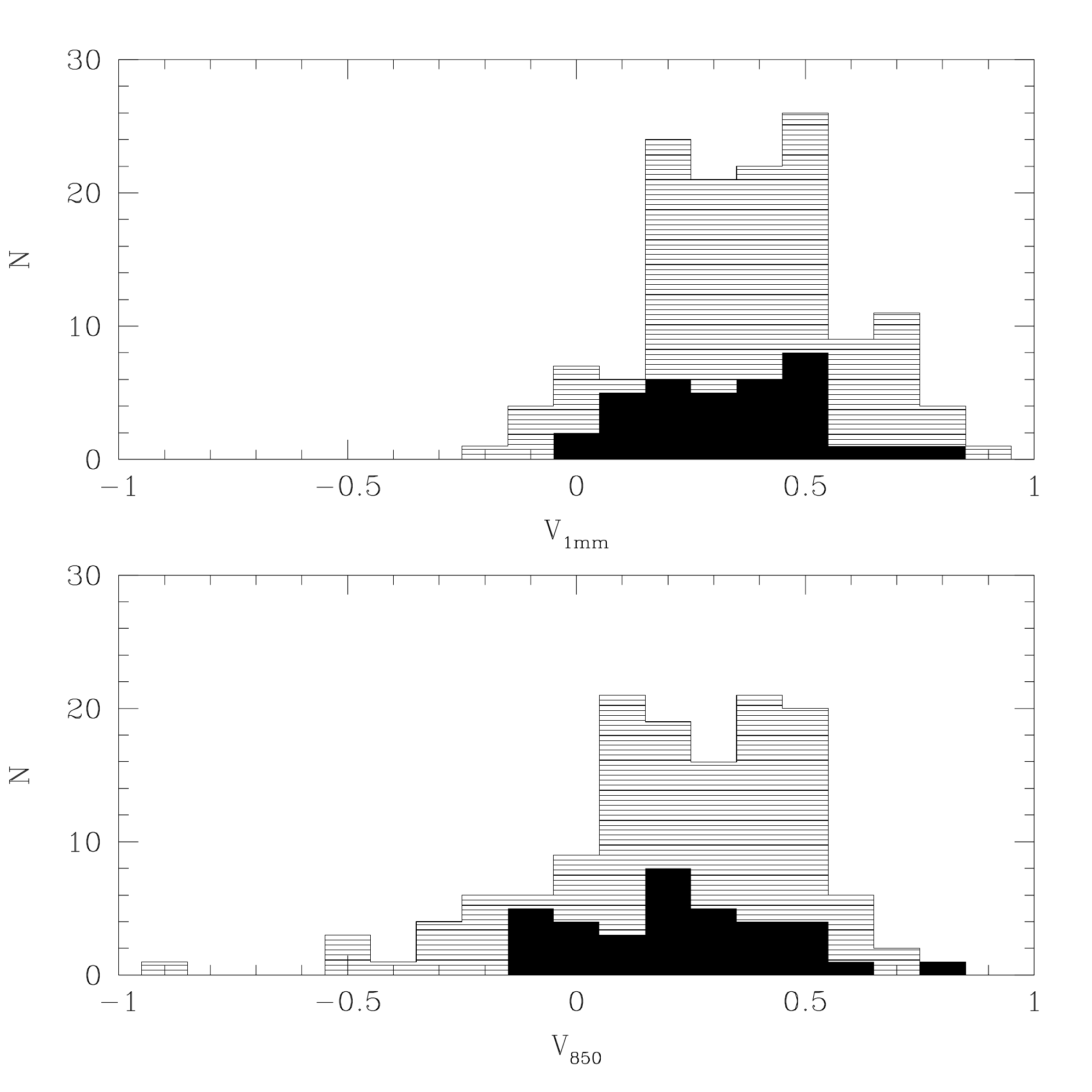}
\caption{Variability index for our sample (top: 1mm, bottom: 850$\mu$m), with FSRQs as the hatched distribution and BL Lacs as the solid distribution. There is no signiÞcant difference in the class distributions in either band; the ``tail" to  the left is populated by objects with errors larger than the intrinsic variability.}
\end{center}
\end{figure}

We roughly characterize the level of variability of each source using the
variability index from Hovatta et al. \cite{hovatta08}:
\begin{equation} \label{eq:varindex}
 V\, =\, \frac{(F_\mathrm{max} - \sigma_{F_\mathrm{max}}) - (F_\mathrm{min} + \sigma_{F_\mathrm{min}})}{(F_\mathrm{max}  - \sigma_{F_\mathrm{max}}) + (F_\mathrm{min} + \sigma_{F_\mathrm{min}})}
\end{equation}

Figure 2 shows the distribution for the \textit{SMA} blazars. Objects with $V \leq 0$ are typically unsuitable for more detailed
variability analysis for one of two reasons: (1) too few data points
or (2) flux measurement uncertainties on the order of the amplitude of observed
variability.  It is important to note that, due to discrepancies
between the sampling frequency in both bands, the variability indices
for the 850$\mu$m band may be artificially depressed due to the fact
that there are not always corresponding measurements at higher
frequencies during flaring epochs. 

\subsection{First-Order Continuous Autoregression}

We follow the method of Kelly et al. \cite{kelly09}, who model quasar optical light curves as a continuous
time first-order autoregressive process (CAR(1)) in order to extract
characteristic time scales and the amplitude of flux variations.
Although flaring behavior is not typically thought of as an
autoregressive process, we find that the light curves are well-fit by
the models and therefore adopt the method here to study blazar
submillimeter light curves.

The CAR(1) process is described by a stochastic differential equation \citep{kelly09},
\begin{equation} \label{eqn:careq}
dS(t)\, =\, \frac{1}{\tau}S(t)\,dt+\sigma\sqrt{dt}\,\epsilon\,(t)+b\,dt,
\end{equation}
associated with a power spectrum of the form
\begin{equation} \label{eqn:carspec}
P_X(f)\, =\, \frac{2\sigma^2\tau^2}{1+(2\pi\tau f)^2}.
\end{equation}
In equations \ref{eqn:careq} and \ref{eqn:carspec}, $\tau$ is called the
``relaxation time'' of the process $S(t)$ and is identified by the break in
$P_X(f)$. The power spectrum appears flat for timescales longer than this and
falls off as $1/f^2$ for timescales shorter than the characteristic
timescale of the process. 

Taking the logarithm of the blazar light curve (in Jy) to be $S(t)$, we adopt
$\tau$ (in days) as the characteristic timescale of variability, after which the
physical process ``forgets'' about what has happened at time lags of
greater than $\tau$. The two other relevant parameters, $\sigma$
and $\mu = b/a$, are the overall amplitude of variability and the logarithm of
mean value of the light curve, respectively.

In the routine, we construct an autoregressive model for the light curves for a minimum of 100,000 iterations and
calculate the value of $\tau$ from the break in the power spectrum in
each instance. Due to the limited number of observations in the 850$\mu$m band, we
performed this autoregressive analysis only for the 1mm light curves,
which typically have more than 10 points per light curve.

\begin{figure}[t]
\begin{center}
\includegraphics[width=0.50\textwidth]{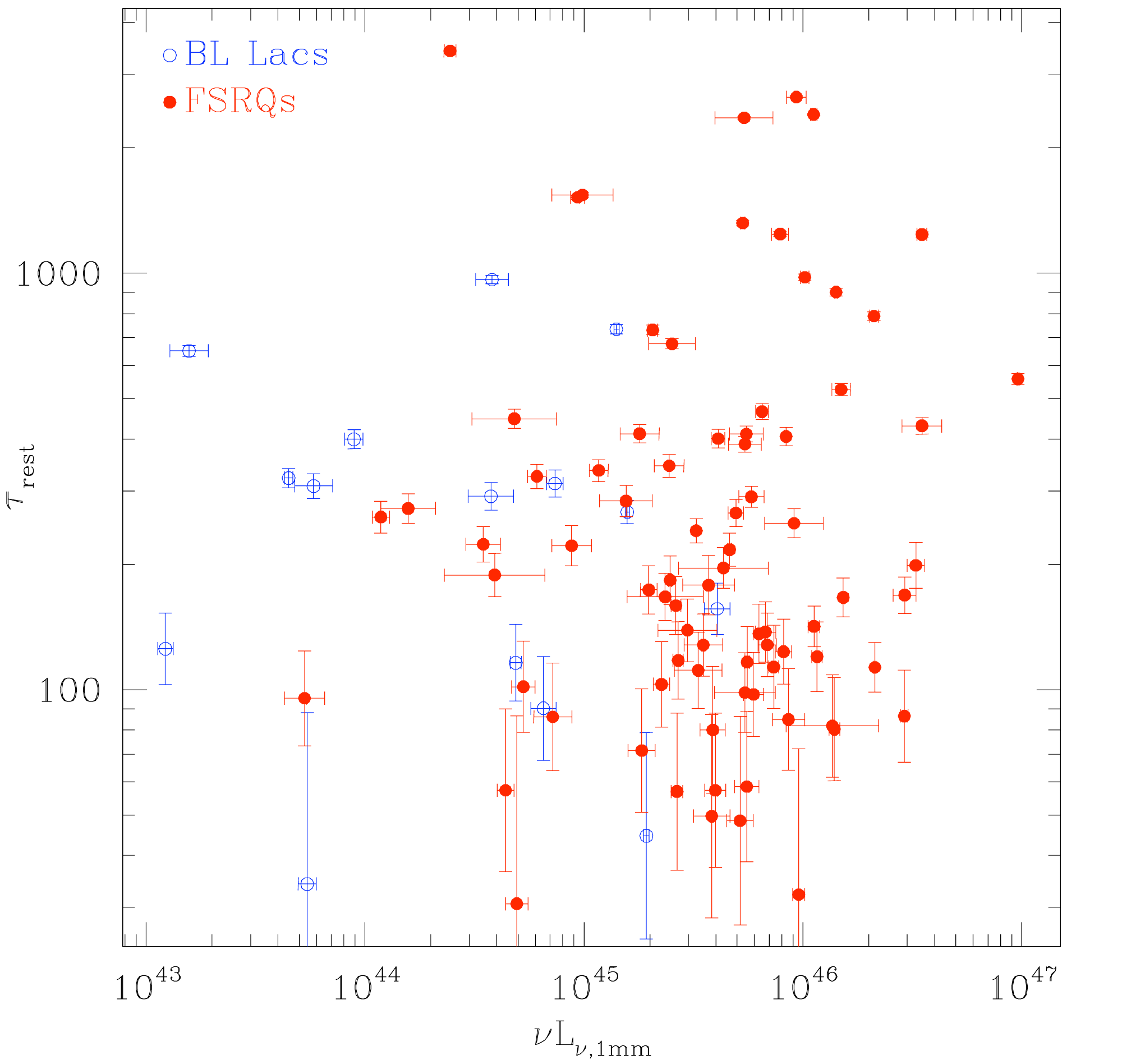}
\caption{Characteristic timescale (days) versus submillimeter luminosity (erg s$^{-1}$) in
  the 1mm band for all objects.  Physically, $\tau$ represents a ``relaxation timescale", the timescale beyond which events are no longer correlated.}
\end{center}
\end{figure}

This method yielded some surprising results. In Figure 3, we see that the BL Lacs and FSRQs exhibit virtually no difference in characteristic timescale, with both classes extending across a large range in $\tau$. Because of the uncertainty for objects with shorter characteristic timescales, it is hard to draw any definitive conclusions about the differences between classes. It is important to note that $\tau$ does not necessarily represent a flaring timescale, which is a behavior that typically operates on a scale of $\sim$10--100 days and not on the longer timescales we see in $\tau$.

\section{CONNECTION WITH GAMMA-RAYS}

\begin{figure*}[t]
\begin{center}
\includegraphics[width=0.65\textwidth]{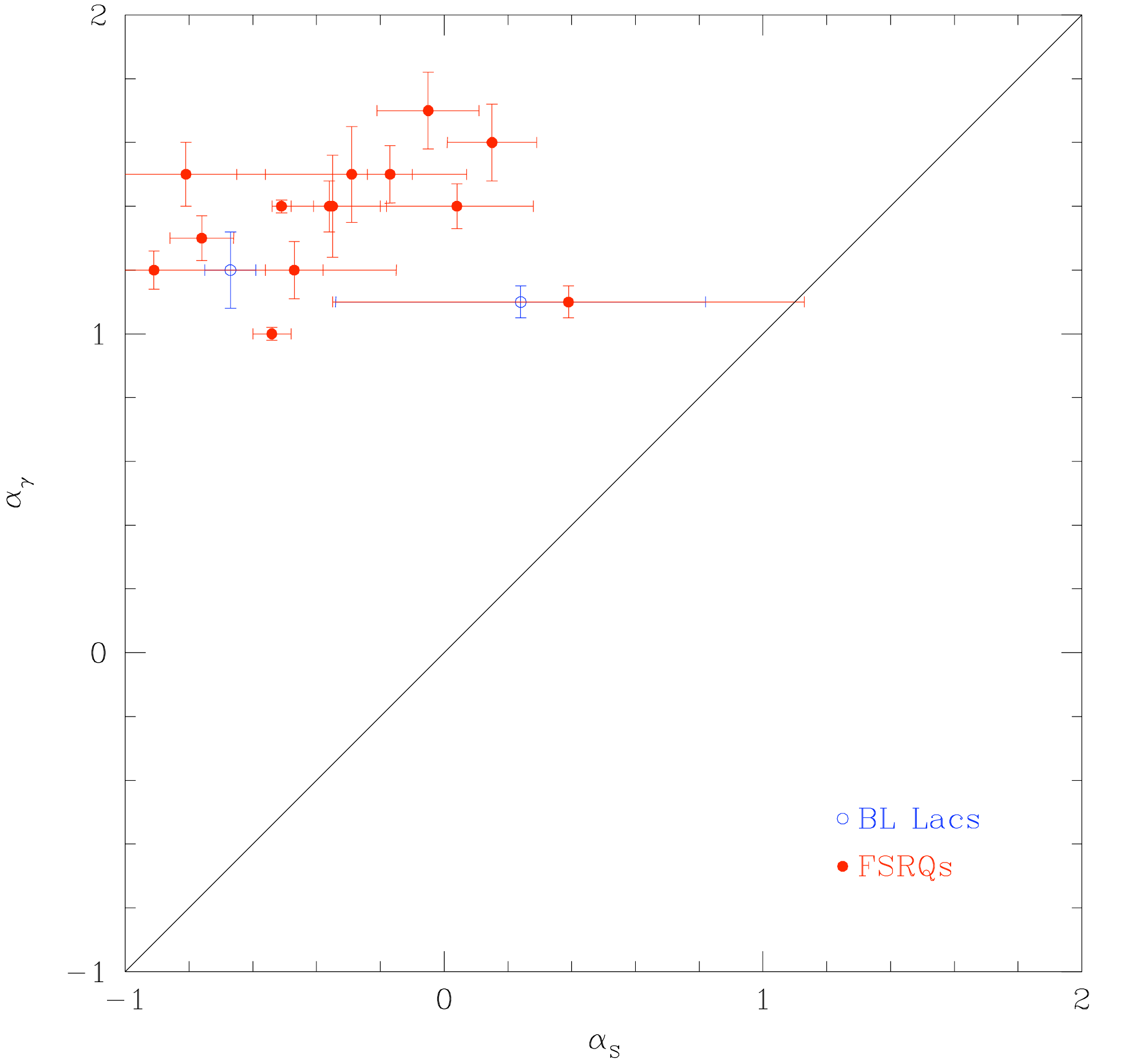}
\label{fig:indexindex}
\caption{The $\gamma$-ray index versus submillimeter index plane. The blazars fall
  more steeply in the $\gamma$-rays than in the submillimeter
  band, where most are, in fact, rising. This LAT-detected sample
  contrasts with the full \textit{SMA} sample, where the blazars are
  more distributed around $\alpha_{\rm S} \sim$ 0.}
\end{center}
\end{figure*}

\begin{figure*}[t]
\begin{center}
\includegraphics[width=0.65\textwidth]{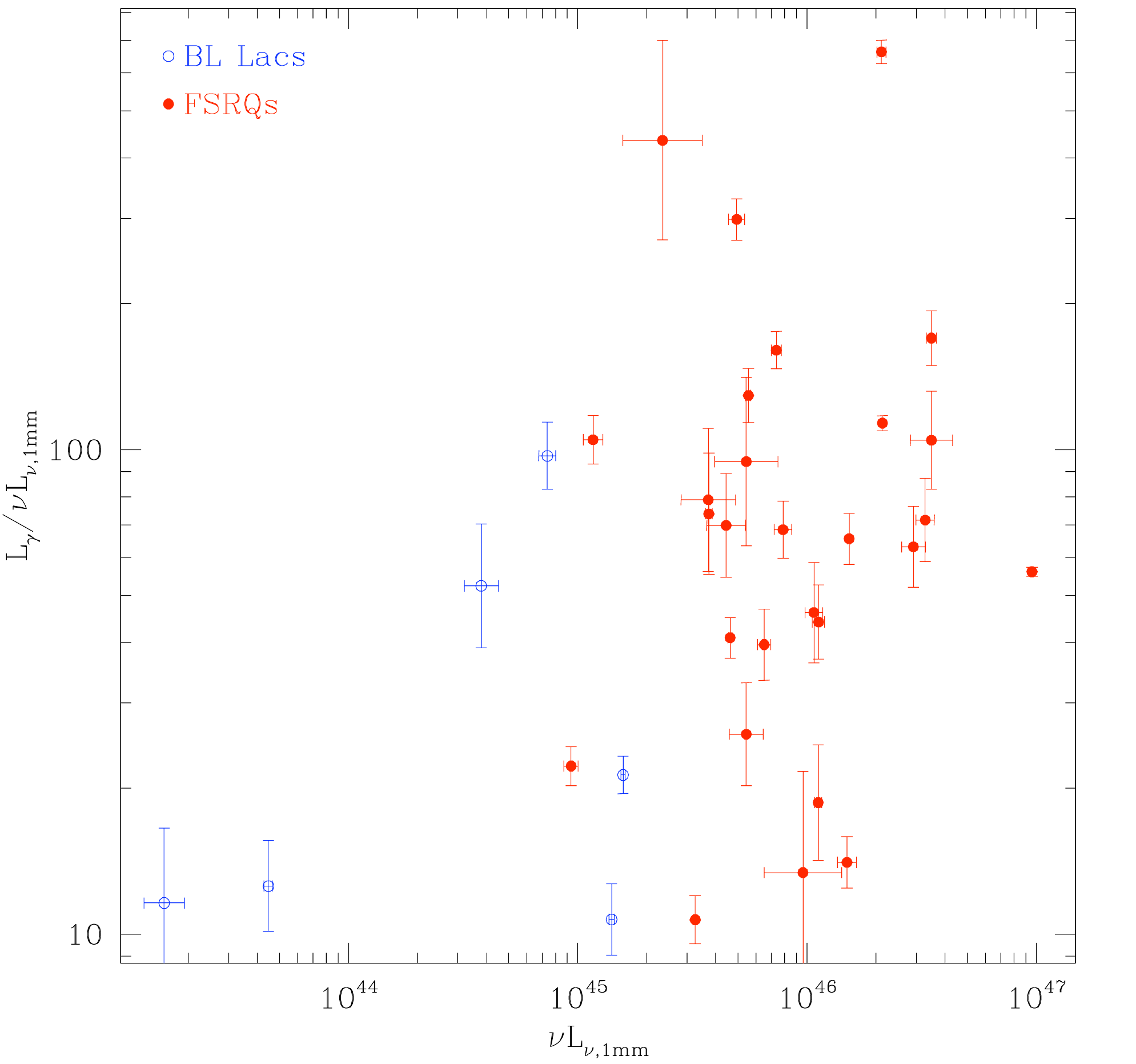}
\label{fig:ratiolum}
\caption{Ratio of $\gamma$-ray luminosity to submillimeter luminosity
  in the 1mm band. The location of an object in this plot should be directly
  correlated with its blazar ``state'', with FSRQs occupying the upper
  right and BL Lacs the lower left. Flat-spectrum radio quasar 3C 454.3 is the object with the highest submillimeter luminosity in this plot.}
\end{center}
\end{figure*}

In general, we find that in the submillimeter, we are observing these
blazars at or near the peak of the synchrotron component ($\alpha_{\rm
S} \sim 0$), but that \textit{Fermi}-detected sources have more
negative energy spectral indices overall than \textit{Fermi}-nondetected
sources. In Figure 4, we see that while the majority of \textit{Fermi}
blazars are observed on the rising part of the synchrotron component
(at lower energies than the peak), all of the objects have very
steeply falling $\gamma$-ray energy spectral indexes, putting the
$\gamma$-ray peak at lower energies than the observed
\textit{Fermi} band. Knowing that we are not observing the synchrotron
and $\gamma$-ray components at analagous points in the spectrum may
allow us to better understand the magnetic field in the parsec-scale
jet region and the population of external photons that is being
upscattered to $\gamma$-rays.

In Figure 5, the ratio between $L_\gamma$ and $\nu L_{\nu \rm
  ,1mm}$ reflects the division between BL Lacs and FSRQs as well as
the presence of SSC versus ERC.  Here, we use submillimeter luminosity
as a proxy for jet power, which is correlated with the integrated
luminosity of the synchrotron component. Elevated $\gamma$-ray luminosity with
respect to the synchrotron component (which is often seen in FSRQs)
suggests the upscattering of external photons off the
synchrotron-emitting electrons.  These objects should occupy the upper
right of the ratio/jet power plot, and BL Lacs, which generally exhibit components
with roughly comparable luminosities, should occupy the lower left. It
is clear from the figure, however, that many FSRQs exhibit ratios similar to
those of the BL Lacs and vis versa. 

Sikora et al. \cite{sikora08} report that, during its flaring epochs, 3C 454.3
transitions from  its typical FSRQ state  to a more BL Lac-like state,
where the synchrotron component  emits much more strongly  compared to
the  $\gamma$-ray component than  during its  ``low state''. 3C 454.3,
which  is  the highest  submillimeter luminosity  FSRQ  in our sample,
would  then shift down and to  the right in Figure 5 when it enters a
flaring  period.  For the   first three  months of the  \textit{Fermi}
mission,  3C 454.3  was  not flaring,  which  may explain  its present
location  in Figure 5. The three   objects for which  there  is a type
discrepancy  between CGRaBS and  LBAS  are all FSRQs  (in CGRaBS)  and
exhibit low luminosity ratios and  high luminosity, which suggest they
may be    undergoing  the  same  changes  as   3C  454.3.   A possible
interpretation of the elevated  luminosity ratios observed in some  BL
Lacs   objects is that  there  has been  a   dramatic increase   in
$\gamma$-ray luminosity  due to ERC,  which would not  be reflected in
the synchrotron component.

\section{CONCLUSIONS}

The motivation for observing blazars in the submillimeter is to study
behavior close to the central engine, where the jet material is
presumably still being accelerated. The separate emission processes
that contribute to overall SED may present differently in BL Lacs and
FSRQs, allowing us to understand the similarities and differences
between blazar types. We have investigated these differences between
objects in terms of submillimeter behavior and, in conclusion, find
that

\begin{itemize}
  \item The \textit{SMA} blazars exhibit submillimeter energy spectral
  indexes that follow the spectral sequence interpretation of blazars.
  \item BL Lacs and FSRQs do not exhibit significant differences in
  amplitude of submillimeter variability or characteristic timescale,
  but our sample of BL Lacs may be dominated by high-peaked BL Lacs
  (HBLs), which exhibit observational similarities with FSRQs.  \item
  Blazar submillimeter light curves are consistent with being produced
  by a single process that accounts for both high and low states, with
  characteristic timescales 10 $< \tau_{\rm rest} <$ 500 days.  \item
  The blazars detected by \textit{Fermi} have synchrotron peaks at
  higher frequencies, regardless of submillimeter luminosity.
\item FSRQs exhibit higher ratios of $\gamma$-ray to submillimeter
  luminosity than BL Lacs (Figure 5), but all objects inhabit a region
  of parameter space suggesting transitions between states during
  flaring epochs.
\end{itemize}

As \textit{Fermi} continues to observe fainter sources, the sample of
objects for which we can perform this type of analysis will increase
and provide better limits on our results. To understand the physical
relevance of these results, however, it is important to be able to
distinguish between the difference in variability between BL Lacs and
FSRQs. One avenue for exploring this difference is to monitor changing
submillimeter energy spectral index and the ratio of $\gamma$-ray to
submillimeter luminosity as functions of time. The full meaning of the
results of our autoregressive method is not yet clear, and will
require better-sampled blazar light curves and the comparison between
$\tau_{\rm rest}$ with physical timescales such as the synchrotron
cooling timescale. These analyses would allow us to place constraints
on the processes occurring near the base of the jet in blazars and
further understand the intimate connection between them.

\begin{acknowledgments}

This work was supported in part by the NSF REU and DoD
ASSURE programs under Grant no. 0754568 and by the Smithsonian
Institution. Partial support was also provided by NASA contract NAS8-39073
and NASA grant NNX07AQ55G. We have made use of the SIMBAD database, operated at CDS,
Strasbourg, France, and the NASA/IPAC Extragalactic Database (NED)
which is operated by the JPL, Caltech, under contract with NASA.

\end{acknowledgments}

\bibliography{master}

\end{document}